\documentstyle[11pt,newpasp,twoside,epsf]{article} 
\begin{document}
\title{New Direct Observational Evidence for Kicks in SNe}
\author{T. M. Tauris \& E. P. J. van den Heuvel}

\affil{CHEAF, and Astronomical Institute
       ``Anton Pannekoek'', University of Amsterdam, Kruislaan 403,
       1098 SJ Amsterdam, The Netherlands}
      
\section{Introduction}
We present an updated list of direct strong evidence in favour of kicks being
imparted to newborn neutron stars. In particular we discuss the new
cases of evidence resulting from recent observations of the {X}-ray binary
Circinus~{X}-1 and the newly discovered binary radio pulsar PSR~J1141--6545.
We conclude that the assumption that neutron stars receive a kick velocity 
at their formation is unavoidable (van~den~Heuvel \& van~Paradijs 1997).

This assumption 
explains a large variety of observations, ranging from direct observed
properties of individual binary pulsars and Be/{X}-ray binaries to the
observed birth rates and dynamical properties of the populations of
LMXBs, binary recycled pulsars as well as the motion and distribution
of single pulsars.
Below we give an updated list in favour of kicks based on the compilation
given by van~den~Heuvel \& van~Paradijs (1997) -- see references therein
for details.

\section{List of evidence for asymmetric supernovae and resulting kicks}
\begin{itemize}
\item[$\bullet$] High radial velocity of Circinus {X}-1 (Tauris et al. 1999),
                 cf. Sect.~3 
\vspace{-0.20cm}
\item[$\bullet$] Nonalignment of spin and orbit in PSR~J0045--7319 (Kaspi et al. 1996)
\vspace{-0.20cm}
\item[$\bullet$] Geodetic precession and evolution of PSR 1913+16 (Wex et al. 1999)
\vspace{-0.20cm}
\item[$\bullet$] Low eccentricity of PSR~J1141--6545 (Tauris \& Sennels 2000), 
                 cf. Sect.~4
\vspace{-0.20cm}
\item[$\bullet$] High eccentricities of Be/{X}-ray binaries (Verbunt \& van~den~Heuvel 1995)
\vspace{-0.65cm}
\item[$\bullet$] High velocities of certain BMSP/LMXBs (e.g. Tauris \& Bailes 1996)
\vspace{-0.20cm}
\item[$\bullet$] Population synthesis/incidence arguments (e.g. Dewey \& Cordes 1987)
\end{itemize}
The origin of the kick mechanism is still rather unclear; it either may be
related to a neutrino driven convection instability and
an asymmetric outflow of neutrinos or to mass outflow in a MHD jet during/following 
the core collapse.

\section{Circinus {X}-1 -- survivor of a highly asymmetric SN}
Based on the recently measured (Johnston, Fender \& Wu 1999) radial velocity
of +430~km~s$^{-1}$, Tauris et al. (1999) find that a {\em minimum} 
neutron star kick velocity of $\sim$500~km~s$^{-1}$
is needed to account for such a high system radial velocity
(they find that on {\em average}, a kick of $\sim$740~km~s$^{-1}$ is necessary).
This is by far the largest kick needed to explain the motion of any
observed binary system. It should be noted that this result is independent on the
uncertainty of the exact mass of the companion star (most likely $1 < M_2/M_{\odot} < 2$).

\section{PSR~J1141--6545 -- a young pulsar with an old deg. companion}
The very recently discovered non-recycled binary pulsar
PSR~J1141--6545 (Manchester~et~al.~1999) has a massive companion
($M_2 \ge 1.0\,M_{\odot}$) in an eccentric 0.198 days orbit.
The pulsar's high value of $\dot{P}_{\rm spin}$ ($4.31\times 10^{-15}$)
in combination with its relatively slow rotation rate
($P_{\rm spin}=394$ ms)
and non-circular orbit ($e=0.17$) identifies this pulsar as being
young and the last formed member of a double degenerate system.
Given $P_{\rm orb}=0.198$ days it is evident that a non-degenerate
star can not fit into the orbit without filling its Roche-lobe.
Based on evolutionary considerations and population synthesis,
Tauris \& Sennels (2000) demonstrate that the companion is most likely
to be an {O}-{Ne}-{Mg} white dwarf. In that case the system resembles
PSR~B2303+46: the first neutron star -- white dwarf binary system observed, 
in which the neutron star was born {\em after} the formation of the white dwarf
(van~Kerkwijk \& Kulkarni 1999).
The existence of a minimum
eccentricity for systems undergoing a symmetric SN
follows from celestial mechanics (e.g. Flannery \& van~den~Heuvel 1975):
$e = (M_{\rm 2He}-M_{\rm NS})/(M_{\rm WD}+M_{\rm NS})$\\
If one assumes
$M_{\rm 2He}>M_{\rm He}^{\rm crit}\simeq 2.5\,M_{\odot}$,
$M_{\rm WD}^{\rm max}< 1.4\,M_{\odot}$ and $M_{\rm NS}=1.3\,M_{\odot}$ it follows
that $e>0.45$ (Note, that this result also remains valid
if the companion star should turn out to be another
neutron star with a similar mass).
In order to reproduce the low observed eccentricity ($e=0.17$)
Tauris \& Sennels (2000) conclude that the neutron star must have
received a natal kick velocity at birth of $>$100~km~s$^{-1}$.
The evidence of asymmetry remains even if one assumes an extremely
low mass for the exploding naked helium star (e.g. adopting
$M_{\rm He}^{\rm crit}=2.0\,M_{\odot}$ results in a minimum post-SN
eccentricity of 0.26).

\end{document}